\documentclass[%
 reprint,
 prc,
 amsmath,amssymb,
 aps,
 lengthcheck,
floatfix,
]{revtex4}

\usepackage{graphicx}
\usepackage{dcolumn}
\usepackage{bm}
\usepackage{hyperref}
\usepackage[utf8]{inputenc}
\usepackage{multirow}
\usepackage{soul}
\usepackage{float}
\usepackage{graphicx}
\usepackage{times}
\usepackage[normalem]{ulem}
\usepackage{color}
\usepackage{cellspace}
\usepackage{lipsum}

\newcommand{\be}{\begin{equation}}
\newcommand{\ee}{\end{equation}}
\newcommand{\bea}{\begin{eqnarray}}
\newcommand{\eea}{\end{eqnarray}}
\newcommand {\non}{\nonumber\\}
\newcommand {\nonu}{\nonumber}

\newcommand{\comment}[1]{}
\renewcommand\sout{\bgroup \color{red} \ULdepth=-.5ex \ULset}
\def\simge{\mathrel{\rlap{\raise 0.511ex
     \hbox{$>$}}{\lower 0.511ex \hbox{$\sim$}}}}
\def\simle{\mathrel{\rlap{\raise 0.511ex
      \hbox{$<$}}{\lower 0.511ex \hbox{$\sim$}}}}

\begin{document}


\title{Bayesian reconstruction of nuclear matter parameters from the equation of state of  neutron star matter }

%
\author{Sk Md Adil Imam$^{1,2}$}
\author{N. K. Patra$^3$}
\author{C. Mondal$^4$}
\author{Tuhin \surname{Malik}$^5$}
\author{B. K. \surname{Agrawal}$^{1,2}$}
\email{bijay.agrawal@saha.ac.in}

\affiliation{$^1$Saha Institute of Nuclear Physics, 1/AF 
Bidhannagar, Kolkata 700064, India.}  
\affiliation{$^2$Homi Bhabha National Institute, Anushakti Nagar, Mumbai 400094, India.}  
\affiliation{$^3$Department of Physics, BITS-Pilani, K. K. Birla Goa
Campus, Goa 403726, India}

\affiliation{$^4$Laboratoire de Physique Corpusculaire, CNRS, ENSICAEN,
UMR6534, Université de Caen Normandie, F-14000, Caen Cedex, France}

\affiliation{$^5$CFisUC, Department of Physics, University of Coimbra,
3004-516 Coimbra, Portugal}

\date{\today}

\begin{abstract} 
The nuclear matter parameters (NMPs), those underlie in the construction
of the equation of state (EoS) of neutron star matter, are not directly
accessible. A Bayesian approach is applied to reconstruct the
posterior distributions of NMPs from  EoS of neutron star matter.
The constraints on lower-order parameters as imposed by the finite
nuclei observables are incorporated through appropriately chosen prior
distributions.  The calculations are performed with two sets of pseudo
data on the EoS whose true models are known. The median values of second
or higher order NMPs show sizeable deviations from their true values and
associated uncertainties are also larger. The sources of these uncertainties
are identified as (i) the correlations among various
NMPs and (ii) leeway in the EoS of symmetric nuclear matter,
symmetry energy, and  neutron-proton asymmetry which propagates into the posterior distributions of the NMPs.

\end{abstract}


\maketitle
\section{Introduction}

The bulk properties of neutron stars are instrumental in constraining the
equation of state (EoS) of dense matter  \cite{Webb2007,Lattimer2013}. The
conditions of charge neutrality and $\beta$-equilibrium imposed on
the neutron star matter renders it highly asymmetric leading
to neutron-proton ratio much larger than unity.  The nuclear
part of the EoS can be decomposed into two main components: the EoS
of symmetric nuclear matter (SNM) and density-dependent symmetry
energy. The knowledge of the EoS of neutron star matter may provide an
alternative probe to understand the behavior of underlying 
symmetric nuclear matter and symmetry energy over
a wide range of density which may not be readily accessible in the
terrestrial laboratory. Usually, the components of neutron star matter
EoS are expressed in terms of nuclear matter parameters (NMPs), namely,
the energy per nucleon for symmetric nuclear matter, symmetry
energy and their density derivatives evaluated at the saturation
density ($\rho_0 \simeq 0.16$ fm$^{-3}$).  The lower-order NMPs,  governing
the behavior of neutron star EoS at low densities are determined
by nuclear models calibrated to the bulk properties of finite nuclei  \cite{Lalazissis1997,Piekarewicz2006,Agrawal2005,Klupfel2008,Sulaksono2009}.
The higher-order NMPs are  generally estimated using  observed maximum neutron
star  mass together with radius and tidal deformability corresponding
to the neutron star with canonical mass  ~\cite{Zhang2018, Cai2021,Gil:2021ols}. Such
investigations due to the lack of availability of enough experimental data
or for sake of simplicity are restricted to a small subspace of NMPs.

Gravitational-wave astronomy through the observations of gravitational
wave signals emitted during the merging of binary neutron stars, promises
unprecedented constraints on the EoS of neutron star matter. The tidal
deformability inferred from these gravitational wave events encodes
information about the EoS. For the first time, gravitational wave
event  GW170817 was observed by LIGO-Virgo detector from a low mass
compact binary neutron star (BNS) merger with the total mass of the
system $2.74_{-0.01}^{+0.04}M_\odot$  \cite{Abbott18a, Abbott2019}.  Another gravitational wave signal likely originating from the
coalescence of BNS GW190425 is observed  \cite{Abbott2020}. These two events have already
triggered many theoretical investigations to constrain the EoS of
neutron star matter  \cite{GW170817,Malik2018,De18,Fattoyev2018a,Landry2019,Piekarewicz2019,Malik:2019whk,Biswas2020,Abbott2020,Thi2021}. The upcoming runs of LIGO-Virgo and the Einstein Telescope are expected to observe many more gravitational wave signals emitted from coalescing neutron stars. The mass and radius of neutron stars, observed either in isolation or in binaries, by the Neutron star Interior Composition Explorer  \cite{Watts2016, Miller2019, Riley2019} have offered complementary constraints on the EoS. A sufficiently large number of such observations may be employed to constrain the NMPs directly which underlie in the construction of the EoS of neutron
star matter. Since one needs to estimate simultaneously the values
of about ten  NMPs, investigations along this direction require
computationally efficient statistical tools which allow the evaluation of the
likelihood function for the experimental data that may require appropriate
marginalization.

A Bayesian approach is often applied to analyze gravitational-wave
signals, which involves nearly fifteen parameters, to infer their source
properties \cite{Ashton2019}. It has been  also extended to investigate
the properties of short gamma-ray burst  \cite{Biscoveanu2020a}, neutron
star  \cite{Coughlin2019,HernandezVivanco2019,Biscoveanu2019a},
the formation history of binary compact objects  \cite{Lower2018,Romero-Shaw2019,Ramos-Buades2020,Romero-Shaw2020a,Zevin2020}, population using hierarchical inference  \cite{Abbott18a,Talbot2019} and to test general relativity  \cite{Keitel_2019,Ashton2020,Payne2019,Zhao2019}. Recently, Bayesian approach has become a   useful statistical
tool for parameter estimation in the field of nuclear and
nuclear-astrophysics. It allows one to obtain joint posterior distributions of the model parameters and the correlations among them for a given set of data.  Various
constraints on the parameters known $\it {a }$ $\it{priori}$ are incorporated
through their prior distributions. 

Extraction of nuclear 
matter properties from the chiral effective field theory (EFT), in particular, the issue of over-fitting by appropriately choosing the prior is described in great detail in Ref. \cite{Wesolowski2016}.  Recently, Bayesian techniques have also been employed to constrain symmetry energy \cite{Somasundaram2021}, masses and radii of neutron stars \cite{Drischler2021} using the bounds obtained from chiral EFT. To obtain the symmetry energy parameters from lower bound on neutron matter-energy \cite{Tews2017}; to extract the crustal properties of neutron star \cite{Carreau2019,Thi2021a}; to limit the bounds on cold neutron star matter EoS from observational constraints \cite{Riley2018,Raaijmakers:2019dks,Jiang:2019rcw}; to test the compatibility of GW170817 event with multi-physics data \cite{Guven2020,Biswas2021a}; to constrain neutron star matter from existing and upcoming constraints on the gravitational wave and pulsar data \cite{Landry2020a}; to limit the neutron star EoS with microscopic and macroscopic collisions \cite{Huth2021}; to filter models based on astrophysical observations \cite{Biswas2021b}; or to limit the reach of nucleonic hypothesis in the astrophysical context \cite{Thi2021,Mondal2021} Bayesian techniques have been used extensively.

It is common to study different correlations in the posterior involving parameters and observables alike, within a Bayesian analysis \cite{Biswas2020,Guven2020,phdthesis,Carreau2019}. The origin of these uncertainties is embedded either in the underlying models used or in the variances of the data employed. Existing correlations and uncertainties among the extracted nuclear matter properties from a plethora of nuclear physics data both from the laboratory as well as from the heavens and theoretical calculations at low densities are often left for interpretation \cite{Biswas2020,Guven2020,Carreau2019}. It is also quite useful to test the limit of a certain type of data on physical quantities which are extracted by employing Bayesian analyses. How far the constraints on the static properties of a neutron star can pinpoint the nuclear matter parameters is still a question of great interest. The same applies to the data from heavy-ion collisions which can probe nuclear matter at supra saturation densities \cite{Russotto11,Russotto16,Adamczewski-Musch20}. These studies can be done only in a controlled environment, as the present observations associate large uncertainties on the data. We have tried to do this by employing theoretical modeling to mimic data on neutron star matter, as well as symmetric matter and symmetry energy. 

 We first build the EoS of the neutron star matter by expanding it around symmetric nuclear matter within the quadratic approximation as most commonly employed. The EoS of symmetric nuclear matter and density-dependent symmetry energy, the two main components, are further expanded around saturation density within the Taylor and $\frac{n}{3}$ expansions.   The expansion
coefficients in the former case are the individual NMPs and their linear
combinations in the latter case. A suitable set of NMPs is chosen so
that the resulting neutron star matter EoS is consistent with the
currently observed maximum mass of $\sim 2M_\odot$ and satisfies the
causality condition.  As test cases, we employ these EoSs as pseudo data
in a Bayesian analysis.  The true values of NMPs for the pseudo data are thus known $\it {a}$ $\it{priori}$. The constraints imposed on the lower-order nuclear
matter parameters by the experimental
data, for the bulk properties of finite nuclei, are incorporated through appropriate choice of the prior
distributions. The median values of marginalized posterior distributions
of NMPs and the associated uncertainties on them as obtained for both
the EoS models show similar trends. The inherent nature of the model
responsible for the deviations in the median values of NMPs from their true values and uncertainties on them has
been identified.

The paper is organized as follows, The Taylor
and $\frac{n}{3}$ expansions for the EoS of neutron star matter are
briefly outlined in Sec. \ref{methe}.   a Bayesian approach is also discussed in the same
section. The results for the posterior distributions of NMPs obtained
from the EoS  of symmetric  nuclear matter, density-dependent symmetry
energy  and
the EoS of neutron star matter is presented in Sec. \ref{results}. The main outcomes
of the present investigation are summarized in the last section.

\section{Methodology} 
\label{methe}
The nuclear part of the  energy per nucleon
 for neutron star matter $\varepsilon(\rho, \delta)$ at a given  total
nucleon density,  $\rho$ and asymmetry, $\delta$ can be decomposed into the
energy per nucleon for the SNM, $\varepsilon(\rho,0)$
and the  density-dependent symmetry energy, $J(\rho)$ in the  parabolic approximation as,

\bea
\varepsilon(\rho, \delta) &=&  \varepsilon(\rho,0)+J(\rho)\delta^2
+..., \label{eq:eos}
\eea
where,  $\delta  = \left(\frac{\rho_n -\rho_p}{\rho}\right )$
with $\rho_n$ and  $\rho_p$ being the neutron and  proton densities,
respectively. The value of $\delta$ at a given $\rho$ is determined by the
condition of $\beta$-equilibrium and the charge neutrality. Once $\delta$
is known, the fraction of neutron, proton, electron, muon can be easily evaluated. 
In the following, we expand $\varepsilon(\rho,0)$ and $J(\rho)$ appearing in
Eq. (\ref{eq:eos})  using Taylor and $\frac{n}{3}$ expansions. 
The coefficients of expansion in case of the Taylor correspond to the
individual nuclear matter parameters. In the latter case, they are expressed as linear combinations of the nuclear matter parameters. These EoSs are used as pseudo data in a Bayesian approach to reconstruct the posterior distributions of
nuclear matter parameters.

\subsection{{Taylor's expansion}}

The $\varepsilon(\rho,0)$ and $J(\rho)$ can be
expanded around the saturation density $\rho_0$ as  \cite{Chen:2005ti,Chen:2009wv,Newton:2014iha,Margueron:2017eqc,Margueron:2018eob},

\bea
\varepsilon(\rho, 0)&=&	\sum_{n} \frac{a_n}{n!}\left(\frac{\rho-\rho_0}{3\rho_0}\right )^n, \label{eq:snm_T} \\ 
J(\rho) &=&	\sum_{n} \frac{b_n}{n!}\left(\frac{\rho-\rho_0}{3\rho_0}\right )^n, \label{eq:sym_T}  
\eea
so that, 
\bea
\varepsilon(\rho, \delta)&=&	\sum_{n} \frac{1}{n!}(a_n + b_n\delta^2)\left(\frac{\rho-\rho_0}{3\rho_0}\right )^n ,\label{eq:ebeta_T} 
\eea
where the coefficients $a_n$ and $b_n$ are the nuclear matter
parameters. We truncate the sum in Eqs. (\ref{eq:snm_T}) and
(\ref{eq:sym_T}) at 4th order, i.e., $n = 0$ - $4$.  Therefore, the
coefficients $a_n$ and $b_n$ correspond to,

\bea
a_n & \equiv & \varepsilon_0, 0, K_0, Q_0, Z_0\label{eq:anm1}, \\
b_n  &\equiv & J_0, L_0, K_{\rm sym,0},Q_{\rm sym,0}, Z_{\rm sym,0}. \label{eq:bnm1}
\eea
In Eqs. (\ref{eq:anm1}) and (\ref{eq:bnm1}), $\varepsilon_0 $ is the binding
energy per nucleon, $K_0$ {the} incompressibility coefficient, $J_0$ {the} symmetry
energy coefficient, it's slope parameter $L_0$, $K_{\rm sym,0}$ {the} symmetry
energy curvature parameter, $Q_0 (Q_{\rm sym,0})$ and $Z_0 (Z_{\rm sym,0})$ 
are related to third and fourth order density derivatives of  $\varepsilon (\rho,0) $ ( $J(\rho)$), respectively. The subscript zero indicates that all
the NMPs are calculated at the saturation density.

It may be noticed from Eq. (\ref{eq:ebeta_T}) that the coefficients
$a_n$ and $b_n$ may display some correlations among themselves provided
the asymmetry parameter depends weakly on the density.  Further, the
Eq. (\ref{eq:ebeta_T}) may converge slowly at high densities, i.e.,
$\rho >> 4\rho_0$. This situation is encountered for the heavier
neutron stars. The neutron stars with a mass around $2M_\odot$, typically
have central densities $\sim 4-6\rho_0$.

\subsection{ ${\frac{n}{3}}$ expansion}
Alternative expansion of $\varepsilon(\rho,\delta)$  can be
obtained by expanding $\varepsilon(\rho,0)$ and $J(\rho)$  as
\cite{Lattimer2015,Gil2017},

\bea
\varepsilon(\rho,0) &=& \sum_{n=2}^6 (a'_{n-2}) \left(\frac{\rho }{\rho_0}\right )^{\frac{n}{3}},\label{eq:snm_n3}\\
J(\rho) &=& \sum_{n=2}^6 (b'_{n-2}) \left(\frac{\rho }{\rho_0}\right )^{\frac{n}{3}},\label{eq:sym_n3}\\
\varepsilon(\rho, \delta) &=& \sum_{n=2}^6 (a'_{n-2} +b'_{n-2} \delta^2 ) \left(\frac{\rho }{\rho_0}\right )^{\frac{n}{3}}.\label{eq:ebeta_n3}
\eea
 We refer {this} as {the} ${\frac{n}{3}}$ expansion. It
is now evident from Eqs.(\ref{eq:snm_n3}) and (\ref{eq:sym_n3}) that the
coefficients of expansion are no-longer the individual nuclear matter parameters unlike in case of Taylor's expansion.
The values of the NMPs can be expressed
in terms of the expansion coefficients $a'$ and $b'$ as,

\bea
\left (
\begin{matrix}
\varepsilon_0\\
0\\
K_0 \\
Q_0\\
Z_0\\
\end{matrix}
\right ) &=& 
\left (
\begin{matrix}
1 &1 &1 & 1&1\\
2 &3 &4 & 5&6\\
-2  &  0 & 4 & 10&18\\
8  &  0 & -8 & -10 & 0\\
-56 & 0 & 40 & 40 & 0\\
\end{matrix}
\right )
\left (
\begin{matrix}
a'_0\\
a'_1\\
a'_2\\
a'_3\\
a'_4\\
\end{matrix}
\right ),\label{mat_a}\\
\left (
\begin{matrix}
J_0\\
L_0 \\
K_{\rm sym,0}\\
Q_{\rm sym,0}\\
Z_{\rm sym,0}\\
\end{matrix}
\right ) &=& 
\left (
\begin{matrix}
1 &1 &1 & 1&1\\
2  &  3 & 4 & 5&6\\
-2  &  0 & 4 & 10&18\\
8 & 0 & -8 & -10 & 0\\
-56 & 0 & 40 & 40 & 0\\ 

\end{matrix}
\right )
\left (
\begin{matrix}
b'_0\\
b'_1\\
b'_2\\
b'_3\\
b'_4\\
\end{matrix}
\right ).\label{mat_b}
\eea
The relations between the expansion coefficients and the NMPs are governed by the nature of functional form for $\varepsilon(\rho,0)$ and $J(\rho)$. The off-diagonal elements in the above matrices would vanish for the Taylor's expansion of  $\varepsilon(\rho,0)$ and $J(\rho)$ as given by Eqs.  (\ref{eq:snm_T}) and (\ref{eq:sym_T}), respectively. Therefore, each of the expansion coefficients are simply the individual NMPs given by Eqs.  (\ref{eq:anm1}) and (\ref{eq:bnm1}). Inverting the matrices in Eqs.  (\ref{mat_a}) and (\ref{mat_b}) we have,
\bea
a'_0&=& \frac{1}{24}(360 \varepsilon_0 + 20K_0   + Z_0),\nonu\\
a'_1&=& \frac{1}{24}(-960 \varepsilon_0 - 56K_0  - 4Q_0 - 4Z_0),\nonu\\
a'_2&=& \frac{1}{24}( 1080\varepsilon_0 + 60K_0 + 12Q_0 + 6Z_0),\nonu\\
a'_3&=& \frac{1}{24}(-576\varepsilon_0 - 32K_0 - 12Q_0 - 4Z_0),\nonu\\
a'_4&=& \frac{1}{24}( 120\varepsilon_0 + 8K_0 + 4Q_0  + Z_0),\label{eq:anm2}\\
b'_0&=& \frac{1}{24}(360J_0 -120L_0  + 20K_{\rm {sym,0}} + Z_{\rm {sym,0}}),\nonu\\
b'_1 &=& \frac{1}{24}(-960J_0 + 328L_0 - 56K_{\rm {sym,0}} - 4Q_{\rm {sym,0}}\nonumber  \\
&&    -4Z_{\rm {sym,0}}),\non
b'_2 &=& \frac{1}{24}(1080J_0 - 360L_0 + 60K_{\rm {sym,0}} + 12Q_{\rm {sym,0}} \nonumber \\
&&   + 6Z_{\rm {sym,0}}),  \non
b'_3 &=& \frac{1}{24}(-576J_0 + 192L_0 - 32K_{\rm {sym,0}} - 12Q_{\rm {sym,0}}  \nonumber \\
&&   -4Z_{\rm {sym,0}}),\non
b'_4 &=& \frac{1}{24}(120J_0 - 40L_0 + 8K_{\rm {sym,0}}  + 4Q_{\rm {sym,0}}  \nonumber \\
&&   + Z_{\rm {sym,0}}).\label{eq:bnm2}
\eea
Each of the coefficients $a'$ and $b'$  are the linear combinations of
nuclear matter parameters in such a way that the lower order parameters
may contribute dominantly at low densities. The effects of higher-order parameters become prominent with the increase in density.

\subsection{Bayesian estimation of NMPs}\label{BA}

A Bayesian approach enables one to carry out detailed statistical analysis
of the parameters of a model for a given set of fit data. It yields
joint posterior distributions of model parameters which can be used to study not only the distributions of given parameters but also to examine correlations among model parameters. One
can also incorporate prior knowledge of the model parameters and various
constraints on them through the prior distributions. This approach  is
mainly based on the Bayes theorem which states  that \cite{Gelman2013},

\begin{equation}
P(\bm{\theta} |D ) =\frac{{\mathcal L } (D|\bm{\theta}) P(\bm {\theta })}{\mathcal Z},\label{eq:bt}
\end{equation}
where $\bm{\theta}$  and $D$ denote the set of model parameters and
the fit data. The $P(\bm{\theta} |D )$ is the joint posterior distribution
of the parameters, $\mathcal L (D|\bm{\theta})$ is the likelihood function, $
P(\bm {\theta })$ is the prior for the  model parameters and $\mathcal Z$
is the evidence. The posterior distribution  of a given parameter can be
obtained by marginalizing $P(\bm{\theta} |D )$ over remaining parameters.
The marginalized posterior distribution for a  parameter $\theta_i$
can be obtained as,

\begin{equation}
 P (\theta_i |D) = \int P(\bm {\theta} |D) \prod_{k\not= i }d\theta_k. \label{eq:mpd}
\end{equation}
We use Gaussian likelihood function defined as, 
\bea
{\mathcal L} (D|\bm{\theta})&=&\prod_{j} 
\frac{1}{\sqrt{2\pi\sigma_{j}^2}}e^{-\frac{1}{2}\left(\frac{d_{j}-m_{j}(\bm{\theta)}}{\sigma_{j}}\right)^2}. 
\label{eq:likelihood}  
\eea
Here the index $ j$ runs over all the data, $d_j$ and $m_j$ are the data
and corresponding model values, respectively.  The $\sigma_j$ are the adopted
uncertainties.  The evidence $\mathcal Z$ in Eq. (\ref{eq:bt}) is obtained
by complete marginalization of the likelihood function. It is relevant when
employed to compare different models. However in the present work $\mathcal
Z$ is not very relevant.  To populate the posterior
distribution of Eq. (\ref{eq:bt}), we implement a nested sampling algorithm
by invoking the Pymultinest nested sampling  \cite{Buchner2014} in the
Bayesian Inference Library  \cite{Ashton2019}.M

\section{Bayesian  reconstruction of NMPs}

\label{results}

We have considered Taylor and ${\frac{n}{3}}$
expansions in the previous section to express the EoS for symmetric
nuclear matter and the density-dependent symmetry energy in terms of
the NMPs.  The EoS for neutron star matter can thus be constructed for a given set of NMPs using Eqs. (\ref{eq:ebeta_T})
and (\ref{eq:ebeta_n3}) in a straightforward way.  On the contrary, it is not evident that how
reliably the values of NMPs can be extracted once the EoS for the
neutron star matter is known. To illustrate, we construct EoS for the
neutron star matter using Taylor and ${\frac{n}{3}}$
expansion for a known set of NMPs. These EoSs are then employed as pseudo
data to reconstruct the marginalized posterior distributions of the
underlying NMPs through a Bayesian approach. Since the true models for the pseudo data are known, the sources of
uncertainties associated with reconstructed NMPs may be analyzed
more or less unambiguously. A significant part of the uncertainties
on the model parameters usually arises from the intrinsic correlations
among them  \cite{Mondal2015,Malik2018}. The
intrinsic correlations among the NMPs are the manifestation of the various constraints
imposed by the fit data  \cite{Cai2021,Carreau2019}. These correlations may also depend on the choice of forms of the functions for the  EoS of symmetric nuclear matter and density-dependent symmetry energy.  

\subsection{ Likelihood function and prior  distributions}

To  obtain the marginalized posterior distributions of model parameters within a
Bayesian approach one simply requires a set of fit data, a theoretical
model, and a set of priors for the model parameters as discussed in Sec.  \ref{BA}.  The likelihood function for
a given set of fit data is evaluated for a  sample of model parameters populated
according to their prior distributions. The joint posterior distributions
of parameters are obtained with the help of the product of the likelihood function and
the prior distributions, Eq. (\ref{eq:bt}).  The posterior distribution for individual
parameters can be obtained by marginalizing the joint posterior distribution with the remaining model
parameters.  If the marginalized posterior distribution of a parameter
is more localized compared to its prior distribution, then, the parameter
is said to be well constrained by the fit data.

\begin{table}[htbp]
\caption{\label{tab1} The values of  nuclear matter parameters (in MeV)
which are employed  to construct  various pseudo data    using the Taylor
and ${\frac{n}{3}}$ expansions.  The parameters $\varepsilon_0$, $K_0$, $Q_0$
and $Z_0$ describes the EoS of the symmetric nuclear matter part and $J_0$, $L_0$,
$K_{\rm sym,0}$, $Q_{\rm sym,0}$ and $Z_{\rm sym,0}$ describes density-dependent symmetry energy. The index 'N' denotes the order of a given NMP. }

\centering
\begin{ruledtabular}
\begin{tabular}{ccccc}
{N} &\multicolumn{2}{c}{Symmetric nuclear matter}
                                  &\multicolumn{2}{c}{Symmetry energy} \\
  \hline
 0& $\varepsilon_0$&-16.0 &  $J_0$& 32.0 \\
 1&                &      &    $L_0$& 50.0\\
 2& $K_0$          & 230  &  $K_{\rm sym,0}$& -100\\ 
 3& $Q_0$          & -400 &  $Q_{\rm sym,0}$& 550\\
 4& $Z_0$          & 1500 &  $Z_{\rm sym,0}$& -750\\
\end{tabular}
\end{ruledtabular}
\end{table}

Our fit data are essentially the pseudo data for the EoS of symmetric
nuclear matter, density-dependent symmetry energy, and the EoS for
neutron star matter constructed from a suitable choice of NMPs as
listed in Table \ref{tab1}.  The values of lower-order  NMPs are close
to those obtained from the SLy4 parameterization of the Skyrme force calibrated to the bulk properties for a few selected finite nuclei  \cite{Chabanat97,Chabanat98}.  The values of second or higher-order
NMPs are modified so that the EoS for the neutron star matter remains
causal for both the Taylor and $\frac{n}{3} $ expansions.
Further, the maximum mass of neutron stars for both the expansions satisfies the current
lower bound of $\sim 2M_\odot$. The likelihood function is obtained using
Eq (\ref{eq:likelihood}) for pseudo data and the corresponding model values with
the standard deviation, $\sigma$ equal to unity at all densities ranging from 0.5 - 6$\rho_0$. The
present investigation may not be sensitive to the choice of the NMPs.

\begin{table}[htbp]
\caption{\label{tab2} 
Two different  sets P1 and P2 for the prior distributions of the nuclear
matter parameters (in MeV). The parameters of  Gaussian  distribution (G) are   $\mu$ (mean) and $\sigma$ (standard deviation).  The parameters
'min' and 'max' denote the minimum and maximum values for the uniform
 distribution (U). The saturation density $\rho_0$ is taken to be  0.16 fm$^{-3}$.} 
   \centering
 \begin{ruledtabular}  
\begin{tabular}{ccccccc}
\multirow{2}{*}{Parameters}&\multicolumn{3}{c}{P1}&\multicolumn{3}{c}{P2}\\ \cline{2-7} \\
&{Pr-Dist}&{$\mu$}&{$\sigma$}&{Pr-Dist. }&{$\mu$}&{$\sigma$}\\ 
 & & {min}&{max} & &{min}&{max}\\
\cline{1-7}
\hline 
{$\varepsilon_0$} & G  & -16 & 0.3  &  G & -16  & 0.3           \\ 

{$K_0$} & G & 240  &  100    &  G &240  &  50            \\ 

{$Q_0$} & U & -2000 & 2000      &  G  & -400 & 400             \\ 

{$Z_0$} & U & -3000 & 3000    &  U   &-3000   & 3000            \\ 
  
{$J_0$} & G  & 32    & 5     &  G    &32  &5             \\ 
  
{$L_0$} & U  & 20    & 150   &    G   &50   &50           \\ 
  

{$K_{\rm sym,0}$} &  U & -1000  &1000     &  G  &-100  &200              \\ 
  

{$Q_{\rm sym,0}$} & U &-2000 & 2000     &  G   &-550&400             \\ 
  

{$Z_{\rm sym,0}$} & U  &-3000 &3000     &  U   &-3000& 3000            \\ 
  
\end{tabular}
\end{ruledtabular}
\end{table}

The calculations are performed for two different sets of priors. In Table
\ref{tab2}, we provide the details for the prior sets P1 and P2. Usually,
if the parameters are known only poorly, their prior distribution is
taken to be uniform.
But, in case,  if some information about a  parameter is known $\it {a}$ $\it {priori}$, one simply
assumes Gaussian distributions for the corresponding parameter. The priors
for $\varepsilon_0$, and $J_0$ are taken to be Gaussian with their means
and standard deviations consistent with the constraints imposed by the
finite nuclei properties. For most of the remaining NMPs, the prior set P1
assumes uniform distributions.   The prior
set P2 further imposes stronger constraints on the lower order parameters
such as $K_0$ and $L_0$ which are consistent with those obtained from
finite nuclei properties. The higher-order parameters are assumed to
have wide Gaussian distributions. We also impose an additional constraint on the symmetry energy so that it always increases monotonically with density.

\subsection{Symmetric nuclear matter  and symmetry energy}

The EoS of the symmetric nuclear matter $\varepsilon(\rho,0)$  and the
density-dependent symmetry energy $J(\rho)$  are the two main components
which govern the symmetric part  and the  deviations from  it in the
EoS of the neutron star matter.  The NMPs which  are required in the
constructions of $\varepsilon(\rho,0)$ are $\varepsilon_0, K_0, Q_0$
and $Z_0$ and those for  $J(\rho)$ are $J_0, L_0, K_{\rm sym,0}, Q_{\rm
sym,0}$ and $Z_{\rm sym,0}$.
The  nuclear matter parameters that appear in the expansions  of
$\varepsilon(\rho,0)$   might be correlated with those NMPs appearing in the expansion of $J(\rho)$ 
(cf Eqs. (\ref{eq:ebeta_T}) and (\ref{eq:ebeta_n3})).  
These correlations
might prevent the NMPs from being determined accurately.  Moreover,  the
accurate values of NMPs may also be masked by the strong correlations of
symmetry energy with the asymmetry parameter $\delta$  which determines
the fractions of different baryons and leptons at a given density.
The sources of uncertainties in the NMPs are intrinsically present in a
EoS model.  To avoid some of these uncertainties, we first consider
a Bayesian estimation of the NMPs for a given $\varepsilon(\rho,0)$
and $J(\rho)$ separately, before embarking on their estimations from a
EoS of the neutron star matter.
\begin{figure}[]
\centering
\includegraphics[width=0.5\textwidth]{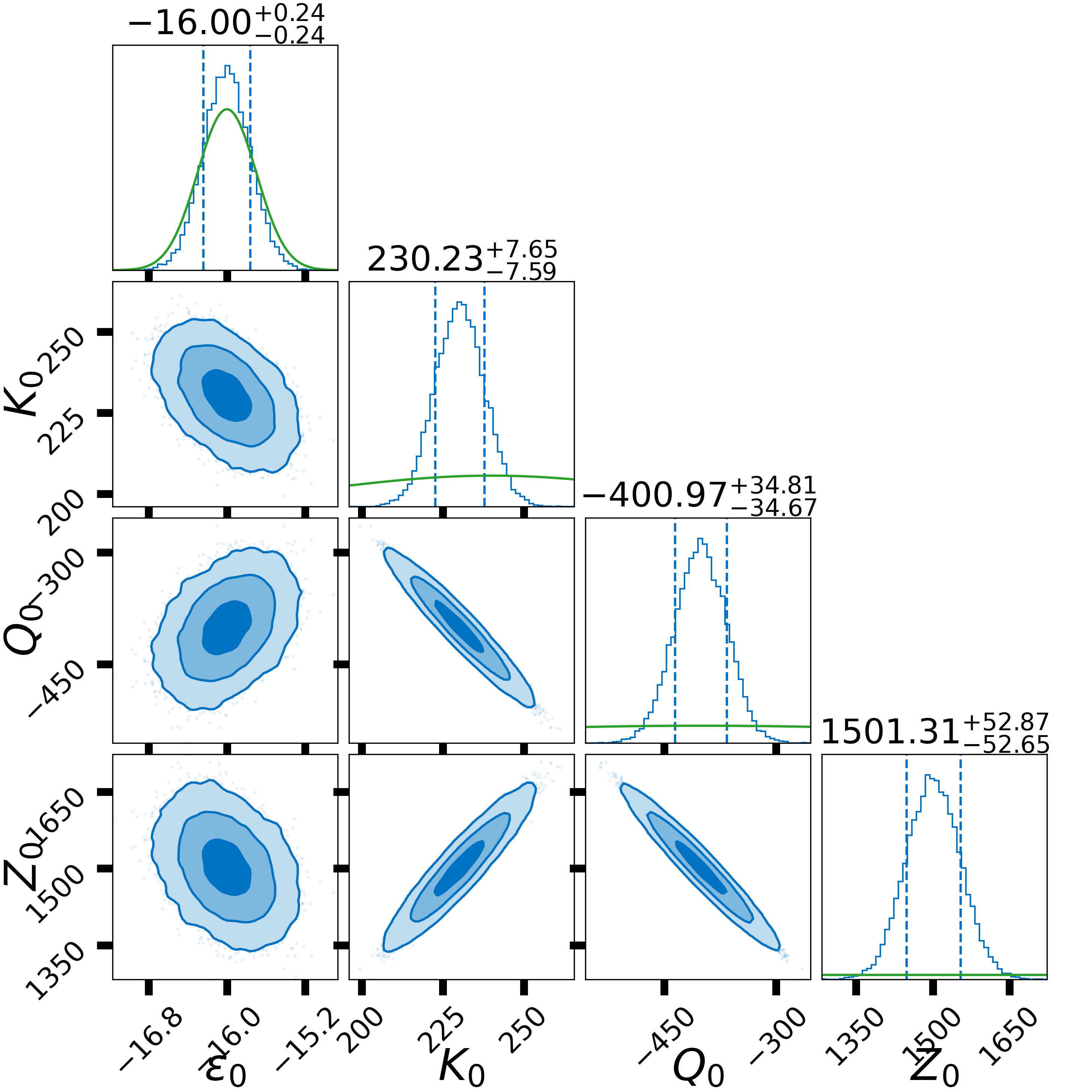}
\vspace{0.5mm}
\includegraphics[width=0.5\textwidth]{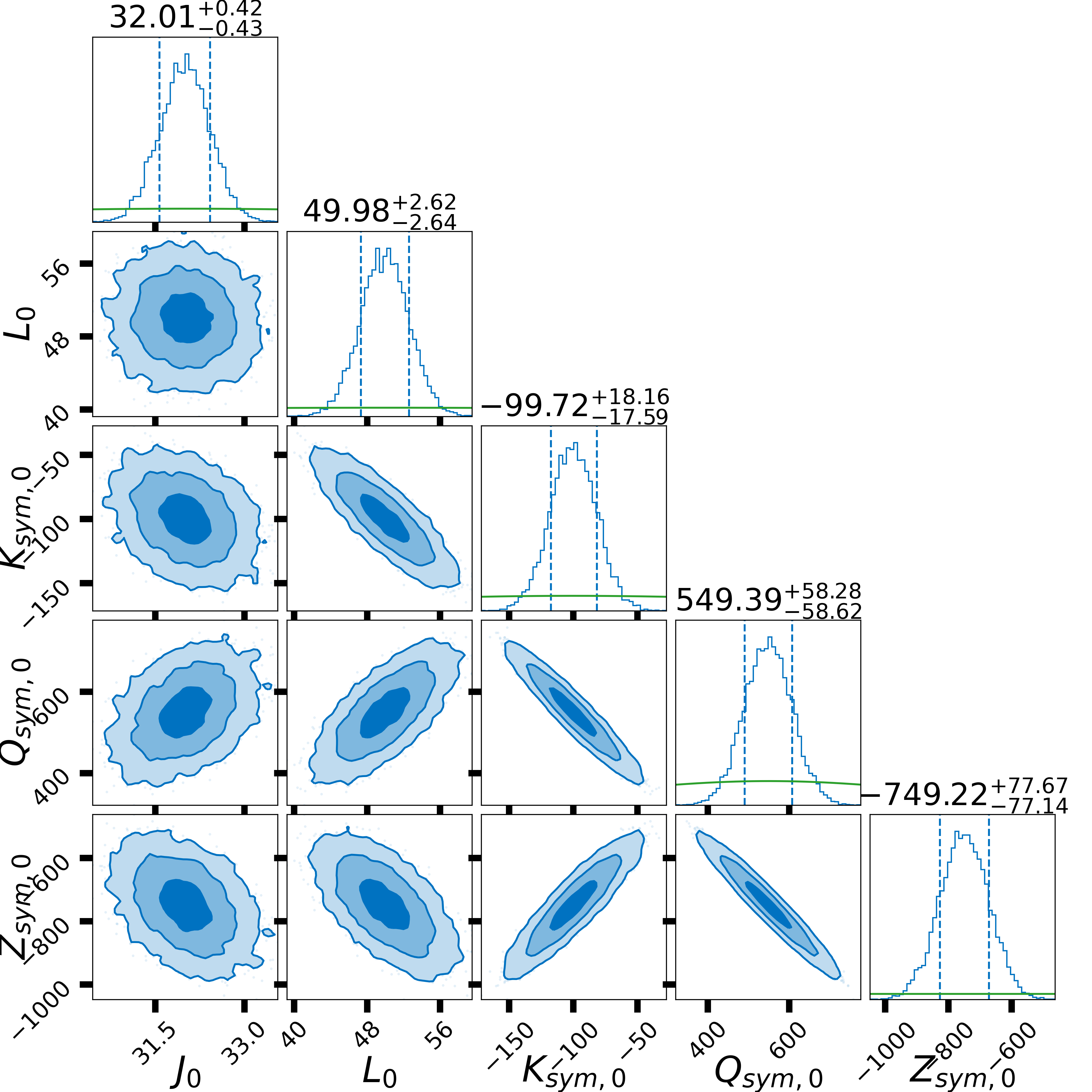}
\caption{Corner plots for marginalized posterior distributions of the
NMPs which appear in the expansion of the EoS for
the symmetric nuclear matter (top)  and those in the density-dependent
symmetry energy (bottom). The results are obtained for the  model
M1  with the prior set P2.  One dimensional posterior distributions
(light blue) plotted along the diagonal plots are also compared with
the corresponding prior distributions (light green). The vertical lines indicate the 68\% confidence interval of the NMPs. The confidence ellipses  for two-dimensional posterior distributions are plotted with $1\sigma$,
$2\sigma$ and $3\sigma$ confidence intervals. \label{fig1} } 
\end{figure}

We perform a Bayesian analysis using Gaussian likelihood
(cf. Eq. (\ref{eq:likelihood}))  which can be easily evaluated for a
set of fit data together with corresponding model values obtained
for a sample of each parameter. We construct two sets of pseudo data
for  $\varepsilon(\rho,0)$ and $J(\rho)$ each. These pseudo data correspond to the Taylor and ${\frac{n}{3}}$ 
expansions referred hereafter as models M1 and M2, respectively.  The values of NMPs used for these pseudo data are the same as listed in Table \ref{tab1}. So, the true values of NMPs for a given pseudo data are known. The marginalized
posterior distributions (PDs) for the NMPs which underlie in the
constructions of $\varepsilon(\rho,0)$ and $J(\rho)$ are obtained
separately.

The median values of NMPs and associated  $1\sigma$ uncertainties from
the marginalized PDs as listed in Table \ref{tab3} are obtained for
the models M1 and M2 for two different prior sets.  The median values of NMPs obtained 
for all the different cases 
are very close to their true values as listed in
Table \ref{tab1}. The uncertainties on the NMPs obtained for
both the prior sets are quite
similar to each other for a given model. However, the uncertainties are
significantly larger for the third and fourth-order NMPs in the case of model M2 in comparison to the ones for M1. The uncertainties for third-order NMPs for M2 are about five times
larger than those for the M1.  It increases to more than ten times for
the fourth-order NMPs.
~                            
The $1\sigma$  uncertainties obtained from the marginalized PDs for NMPs
as listed in Table \ref{tab3} are significantly smaller than the ones
corresponding to their prior distributions. The prior distribution
for a given NMP thus appears relatively uniform compared to its marginalized  PD, as will be seen later.

{\renewcommand{\arraystretch}{1.3}%
\begin{table}[H]
\caption{The median values and the $1\sigma$ errors for the nuclear matter
parameters (in MeV) from their marginalized posterior distributions. The distributions of $\varepsilon_0$, $K_0$, $Q_0$
and $Z_0$ are reconstructed from the EoS of the symmetric nuclear matter  and those for  $J_0$, $L_0$, $K_{\rm sym,0}$, $Q_{\rm sym,0}$ 
and $Z_{\rm sym, 0}$  from the density-dependent symmetry energy. The results are
presented  for the Taylor (M1) and ${\frac{n}{3}}$ (M2) expansions obtained using prior sets P1 and P2.} \label{tab3}
   \centering
 \begin{ruledtabular}  
\begin{tabular}{ccccc}
NMPs & M1-P1 & M1-P2 & M2-P1 & M2-P2\\ [1.5ex]

\hline 
{$\varepsilon_0$} & $ -16.0_{-0.2}^{+0.2} $ & $-16.0_{-0.2}^{+0.2}$ & $-16.0_{-0.2}^{+0.2}$  &  $-16.0_{-0.2}^{+0.2}$   \\[1.5ex]  
{$K_0$} & $230_{-8}^{+8}$ & $230_{-8}^{+8}$  & $230_{-15}^{+14}$    & $230_{-14}^{+13}$ \\[1.5ex] 

{$Q_0$} & $ -402_{-35}^{+35}$ & $-401_{-35}^{+35}$ & $-403_{-124}^{+128}$      &  $-403_{-116}^{+122} $ \\[1.5ex]  
{$Z_0$} &$1502_{-53}^{+53}$ & $1501_{-53}^{+53}$ & $1515_{-773}^{+756}$   & $1517_{-739}^{+711}$ \\[1.5ex]  
  
{$J_0$} & $32.0_{-0.4}^{+0.4}$  &$32.0_{-0.4}^{+0.4} $   & $32.0_{-0.5}^{+0.5}$     &  $32.0_{-0.4}^{+0.4} $  \\[1.5ex]  
  
{$L_0$} &$ 50.0_{-2.8}^{+2.6}$ & $50.0_{-2.6}^{+2.6}$   & $50.0_{-2.6}^{+2.6}$  & $50.0_{-2.5}^{+2.5}$     \\[1.5ex] 
  

{$K_{\rm sym,0}$} & $-100_{-18}^{+18}$ & $-100_{-18}^{+18} $   &  $-100_{-27}^{+27}$ & $-100_{-24}^{+24}$     \\[1.5ex] 
  

{$Q_{\rm sym,0}$} &$551_{-59}^{+58}$ & $549_{-59}^{+58}$ & $548_{-193}^{+184} $   & $551_{-166}^{+163}$    \\[1.5ex]  
  

{$Z_{\rm sym,0}$} &$-750_{-75}^{+80}$ & $-749_{-77}^{+78}$ &$ -734_{-1034}^{+1064}$    & $-759_{-906}^{+936}$    \\[1.5ex] 
  
\end{tabular}
\end{ruledtabular}
\end{table}

\begin{figure}[]
    \centering
    \includegraphics[width=0.5\textwidth]{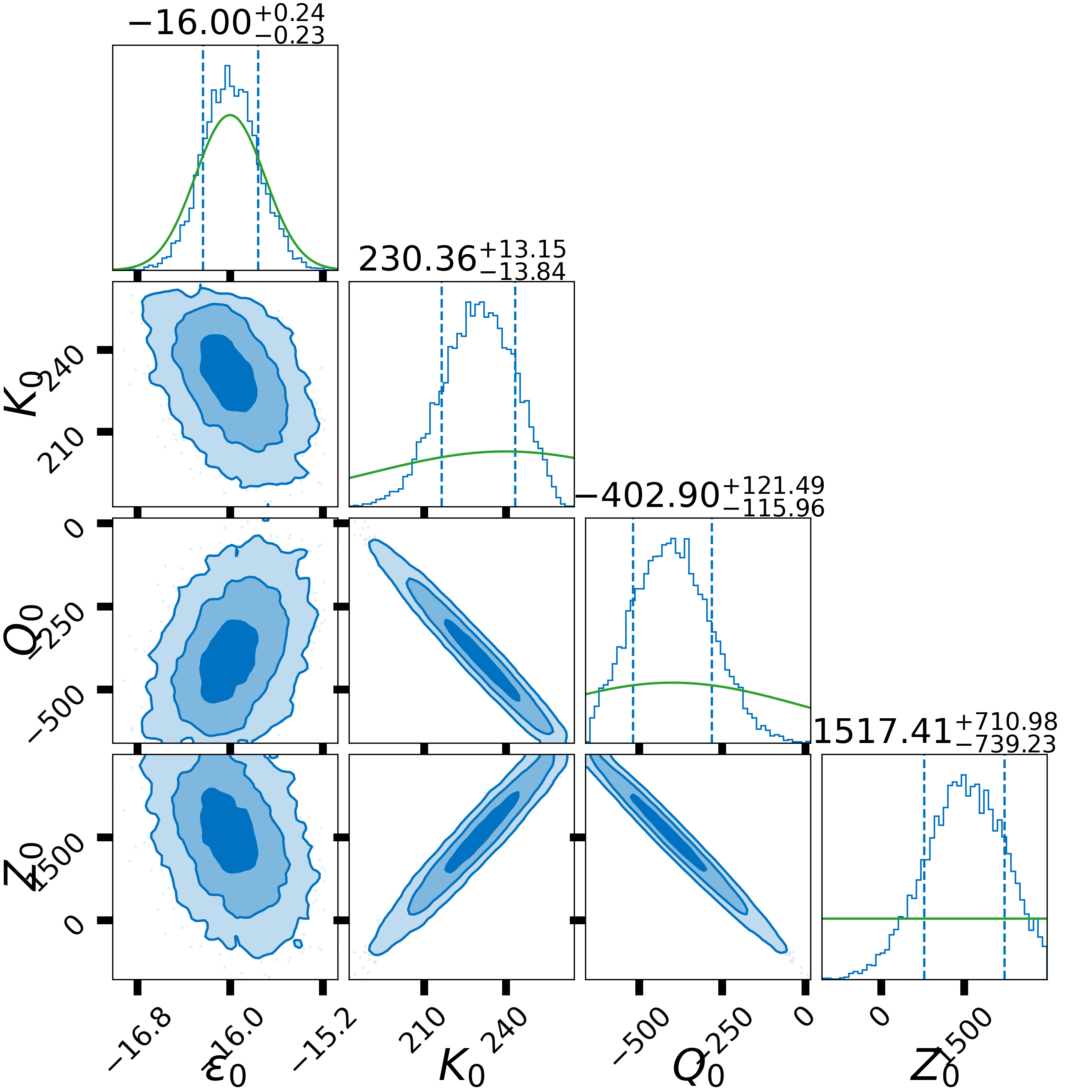}
    \includegraphics[width=0.5\textwidth]{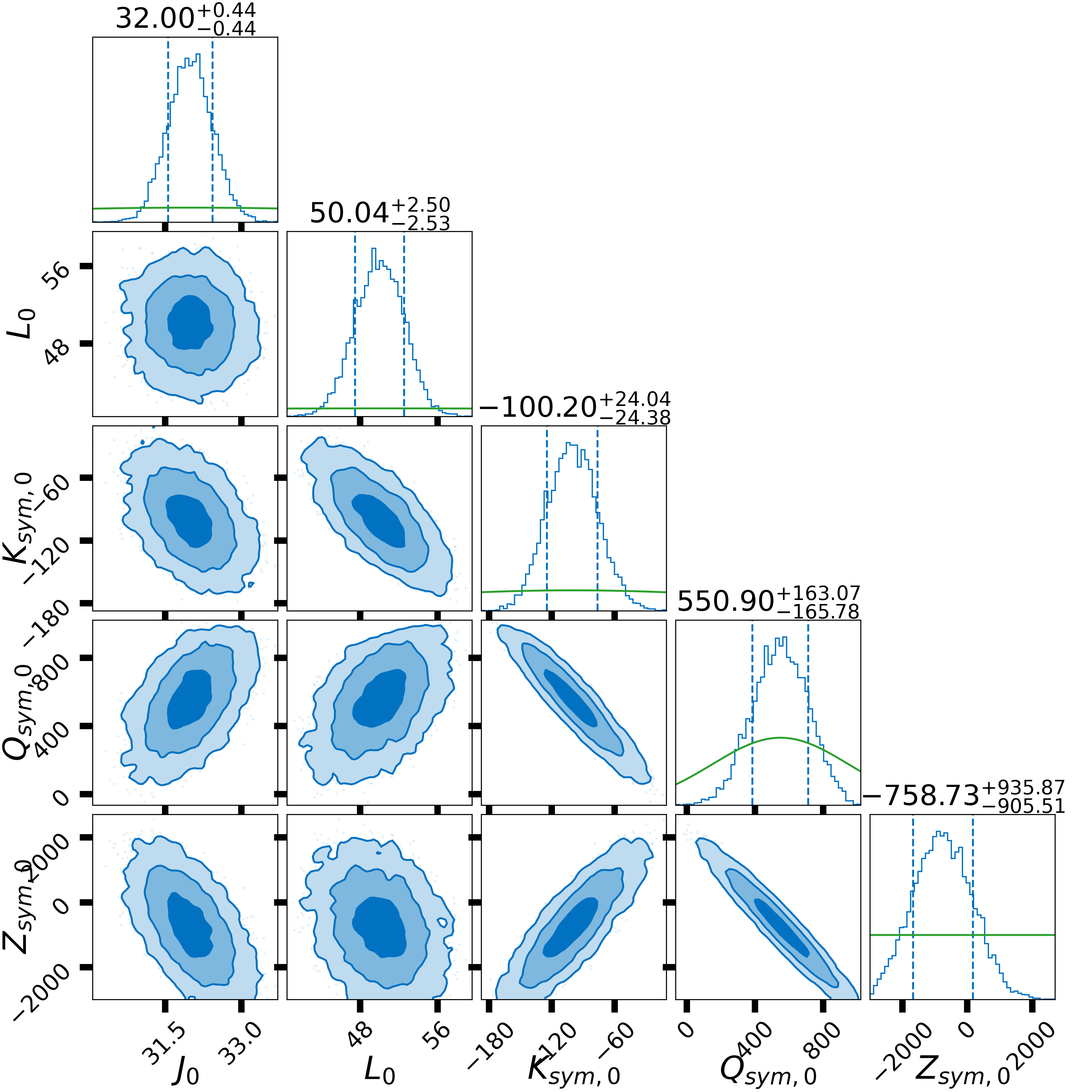}
    \caption{\label{fig2}Same as Fig. \ref{fig1}, but, for model M2.  }
\end{figure}
The marginalized PDs and the confidence ellipses for the NMPs, which
 determine $\varepsilon(\rho,0)$ and $J(\rho)$,   obtained for
models M1 and M2 are displayed as corner plots in Figs. \ref{fig1}
and \ref{fig2}. These results correspond to prior set P2  which assumes very wide Gaussian distributions for the
higher-order NMPs  (see Table
\ref{tab2}). The one-dimensional marginalized  PDs for the NMPs are displayed along
the diagonals of the corner plots (light blue full lines) together
with the corresponding prior distributions ( light green lines). 
These  PDs for all the NMPs are quite symmetric around their median values 
or they represent Gaussian distribution. The PDs for most of the NMPs are localized compared to the corresponding
prior distributions. As a result, the prior distributions for most of the NMPs appear to be flat in comparison to the ones for the marginalized PDs. This is also the reflection of the constraints imposed by the pseudo data. The median values of NMPs are very close to their true values. The confidence ellipses are plotted along with the off-diagonal elements of the corner plots corresponding to $1\sigma$, $2\sigma$, and $3\sigma$ confidence intervals. The width and inclination of the confidence ellipses for a pair of NMPs depend on their covariance which determines the nature of the linear correlations among them  \cite{Reinhard2010,Mondal2015}. It may be noted that the correlation patterns obtained for both models are only marginally different.  However, the uncertainties in the higher-order parameters are significantly larger for the model M2.  This fact may be attributed to some complex intrinsic correlations among the NMPs.It is clear from Eqs. (\ref{eq:anm2}) and (\ref{eq:bnm2}) that the expansion coefficients for the model M2  are the linear combinations of the NMPs unlike those in M1. Furthermore, it can be seen that the higher-order terms in model M2 relative to the lower-order ones have less impact as compared to those in M1. 
\begin{figure}[H]
\centering
\includegraphics[width=0.5\textwidth]{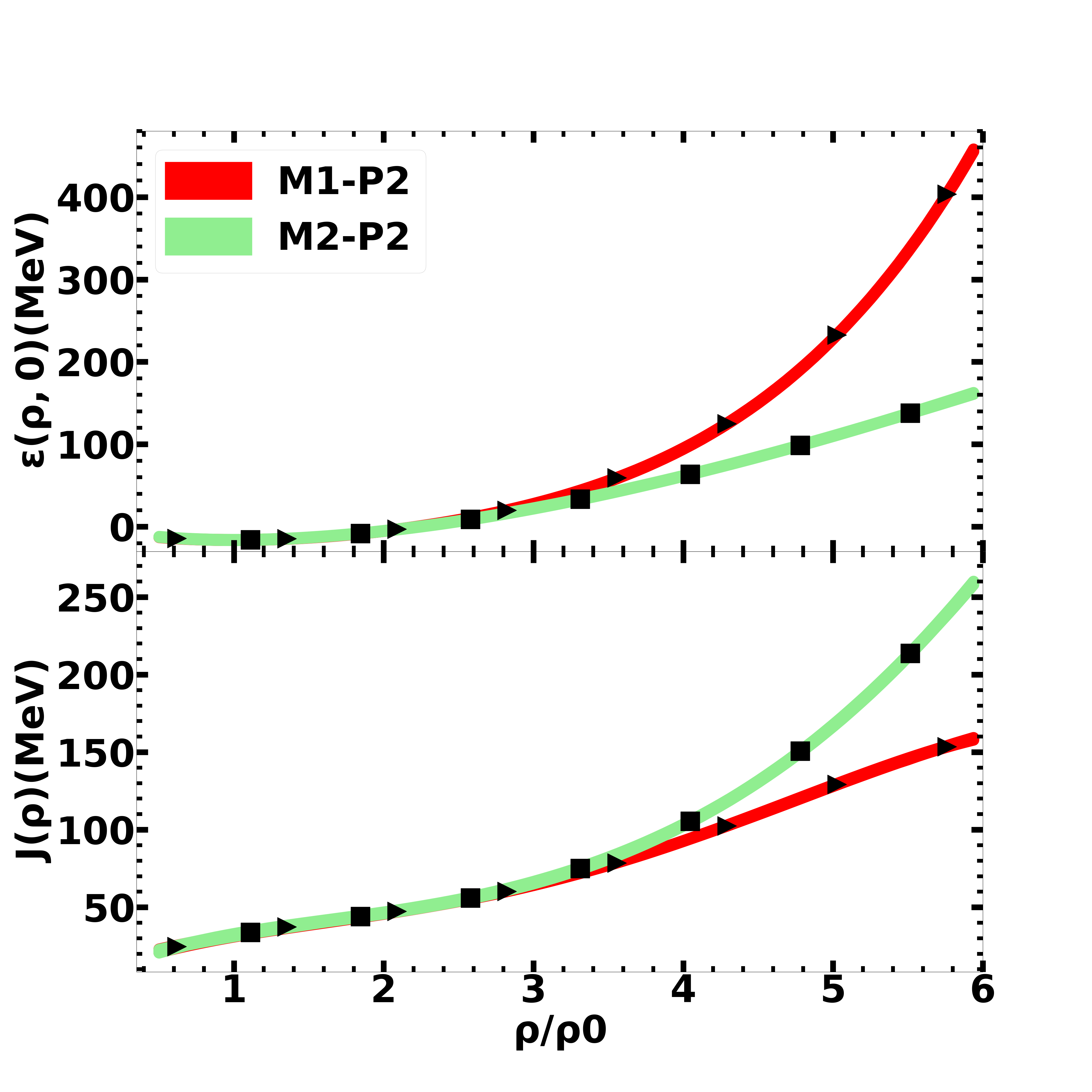}
\caption{\label{fig3}
The  EoS for symmetric nuclear matter (top) and the symmetry energy
(bottom) as a function of density  obtained within 95\% confidence 
 interval from the posterior distributions of nuclear matter parameters for
models M1 and M2 for the prior set P2. The pseudo data for M1 and M2 are shown by triangles and squares, respectively.}
\end{figure}

In Fig.\ref{fig3}, the variations of $\varepsilon(\rho,0)$ and $J(\rho)$
as a function of density with $95\%$ confidence intervals are plotted.
The $95\%$ confidence interval lies in a very narrow range which once
again points to the fact that the large uncertainties on the NMPs are
predominantly due to the correlations among them. The results presented in Figs. \ref{fig1}-\ref{fig3}  provides firm ground to perform the analysis of the NMPs obtained in the following sub-section, using the EoS of neutron star
matter, which involves $\varepsilon(\rho,0)$ and $J(\rho)$, simultaneously.
On passing, we may also remark that though the values of $\varepsilon(\rho,0)$ and $J(\rho)$
for the models, M1 and M2 are obtained using the same set of NMPs, their behavior at high densities is significantly different.  The lower-order NMPs, which govern the low-density behavior of $\varepsilon(\rho,0)$
and $J(\rho)$, maybe model-independent.

\subsection{Neutron star matter}
We now  apply a Bayesian approach to reconstruct the marginalized PDs for the NMPs using
the  EoS for the neutron star matter which satisfies the conditions of
$\beta$-equilibrium and charge neutrality. 
The EoS for neutron star matter $\varepsilon(\rho,\delta)$ can be obtained
using Eq. (\ref{eq:eos}) for a given $\varepsilon(\rho,0)$ and $J(\rho)$. We
construct two sets of pseudo data for $\varepsilon(\rho,\delta)$  corresponding to the models M1 and M2 obtained using  NMPs of  Table \ref{tab1} in Eqs. (\ref{eq:ebeta_T}) and (\ref{eq:ebeta_n3}), respectively.
The marginalized distributions  of all the nine NMPs are reconstructed simultaneously from the pseudo data for $\varepsilon(\rho,\delta)$, since
it consists  of $\varepsilon(\rho,0)$ and $J(\rho)$.  A
Bayesian analysis is performed with models M1 and M2 for prior sets P1
and P2. The median values of the NMPs obtained from the marginalized PDs and the corresponding $1\sigma$
errors are listed in Table \ref{tab4}. The NMPs are somewhat better estimated for the prior set P2. The symmetry energy
slope parameter $L_0$ seems to be a special case as its counterpart in
the symmetric nuclear matter vanishes (cf. Eqs. (\ref{eq:anm1}) and (\ref{eq:bnm1})).  It may be noticed that the uncertainties on $K_{\rm sym,0}$, $Q_{\rm sym,0}$ and
$Z_{\rm sym,0}$ are much larger than their counterparts in the EoS for
symmetric nuclear matter.  The errors on $K_{\rm sym,0}$ are, however,
similar to those derived from bulk properties of finite nuclei or other
correlation systematics  \cite{Mondal:2018lhg,Mondal2017u,Agrawal:2020}, though, we have allowed larger variations of   $K_0$ and $L_0$.  Once,  the sufficiently accurate values of $\varepsilon(\rho,\delta)$  are determined
  from various astrophysical observations,  they can be combined with finite nuclei constraints to obtain $L_0$  and  $K_{\rm sym,0}$ in tighter limits.

The results for  NMPs in Table \ref{tab4}, obtained from the
EoS of neutron star matter are substantially different from those of
Table \ref{tab3}, which were determined separately   from the EoS of
symmetric nuclear matter and the density-dependent symmetry energy.
In general, these  differences   can be summarized as follows, (i)  the median values
of NMPs in Table \ref{tab4} show larger deviations from their true
values compared to those in Table \ref{tab3}, (ii)  the uncertainties
on the NMPs determined  from the EoS of neutron star matter are several
times larger for most of the NMPs, (iii) the uncertainties on the  $Z_0$ and
$Z_{\rm sym,0}$ in Table \ref{tab4} are somewhat asymmetric about their
median values reflecting their non-Gaussian nature,  and (iv) the ratios of uncertainties
between the models, M1 and M2 were obtained for third and fourth-order
NMPs  listed in Table \ref{tab4} are significantly smaller than the ones
 in Table \ref{tab3}. This already provides  us some clue that there are additional sources
of uncertainties on the NMPs determined from the EoS of the neutron
star matter.
{\renewcommand{\arraystretch}{1.3}
\begin{table}[htbp]
\caption{Same as Table \ref{tab3}, but, the posterior distributions for all  the nuclear matter parameters are
reconstructed  simultaneously from  the EoS for the neutron star matter.}
\label{tab4}
   \centering
 \begin{ruledtabular}  
\begin{tabular}{ccccc}
NMPs & M1-P1 & M1-P2 & M2-P1 & M2-P2\\ [1.5ex]
 
\hline 
{$\varepsilon_0$} & $ -16.0_{-0.3}^{+0.3} $ & $-16.0_{-0.3}^{+0.3}$ & $-16.0_{-0.3}^{+0.3}$  &  $-16.0_{-0.3}^{+0.3}$   \\[1.5ex]  
{$K_0$} & $187_{-56}^{+65}$ & $221_{-28}^{+36}$  & $213_{-40}^{+47}$    & $230_{-25}^{+28}$ \\[1.5ex] 

{$Q_0$} & $ -367_{-220}^{+196}$ & $-471_{-123}^{+113}$ & $-327_{-198}^{+243}$      &  $-412_{-123}^{+159} $ \\[1.5ex]  
{$Z_0$} &$1518_{-236}^{+258}$ & $1632_{-157}^{+152}$ & $1307_{-1656}^{+1069}$   & $1637_{-1206}^{+835}$ \\[1.5ex]  
  
{$J_0$} & $31.8_{-2.6}^{+2.5}$  &$32.0_{-2.7}^{+2.6} $   & $32.0_{-2.5}^{+2.5}$     &  $32.0_{-2.4}^{+2.6} $  \\[1.5ex]  
  
{$L_0$} &$ 52.8_{-19}^{+25}$ & $55.5_{-16}^{+17}$    & $53.1_{-19.3}^{+23.8}$  & $51.0_{-13.9}^{+14.0}$     \\[1.5ex] 
  

{$K_{\rm sym,0}$} & $-34_{-178}^{+142}$ & $-108_{-72}^{+76} $   &  $-114_{-138}^{+113}$ & $-106_{-70}^{+70}$     \\[1.5ex] 
  

{$Q_{\rm sym,0}$} &$220_{-563}^{+755}$ & $486_{-264}^{+257}$ & $562_{-488}^{+572} $   & $522_{-241}^{+248}$    \\[1.5ex]  
  

{$Z_{\rm sym,0}$} &$807_{-1527}^{+1341}$ & $100_{-668}^{+876}$ &$ -60_{-1921}^{+1944}$    & $-323_{-1643}^{+1920}$    \\[1.5ex] 
  
\end{tabular}
\end{ruledtabular}
\end{table}
\begin{figure}[]
\centering
\includegraphics[width=0.5\textwidth,height=0.5\textheight]{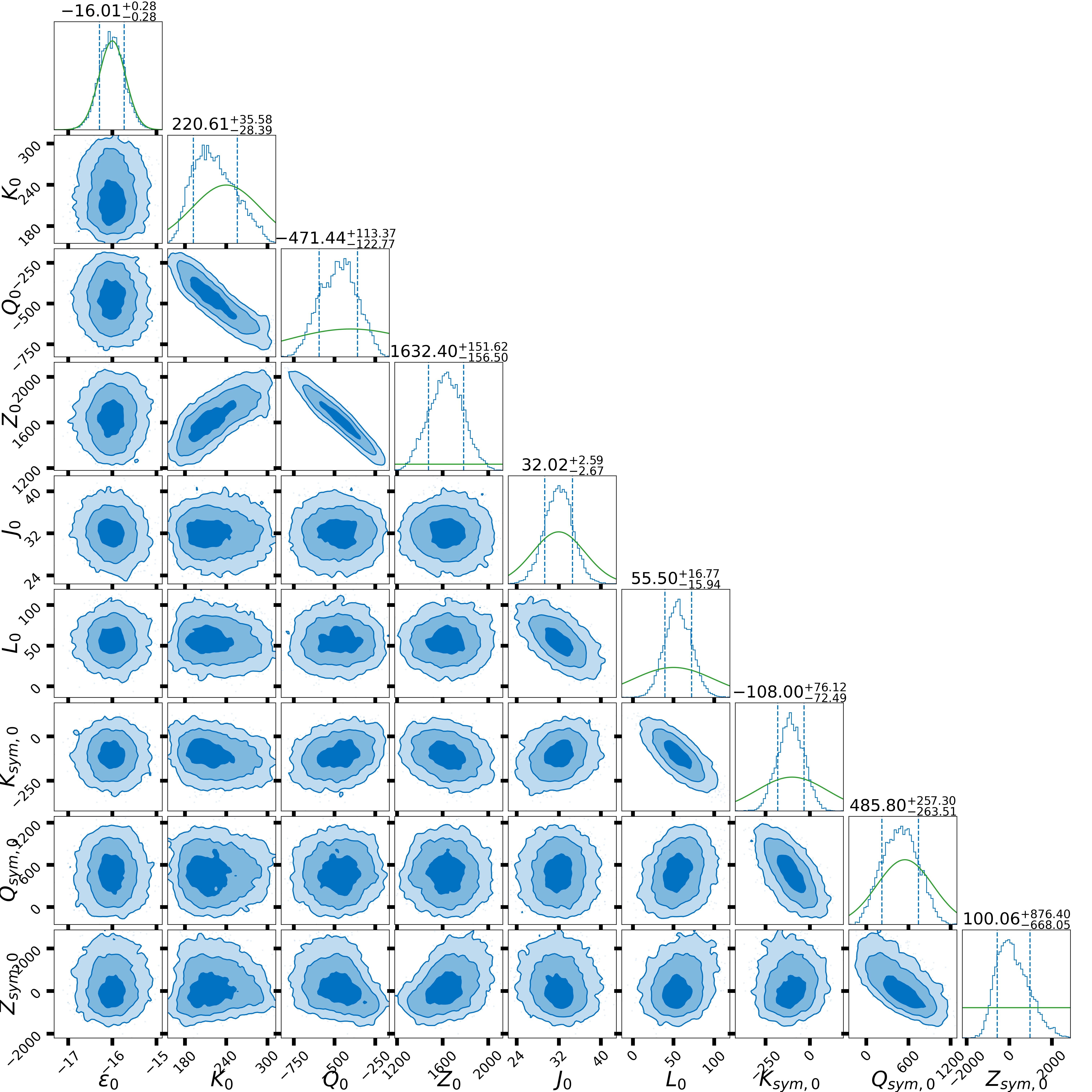}
\caption{Corner plots for the marginalized posterior distributions of  nuclear
matter parameters (in MeV)  obtained from the EoS for the neutron star
matter for the model M1 with prior set P2.  The prior distributions
(light green) are also plotted for the comparison.}
    \label{fig4}
\end{figure}
{\renewcommand{\arraystretch}{1.5}%
\begin{figure}[]
    \centering
    \includegraphics[width=0.5\textwidth,height=0.5\textheight]{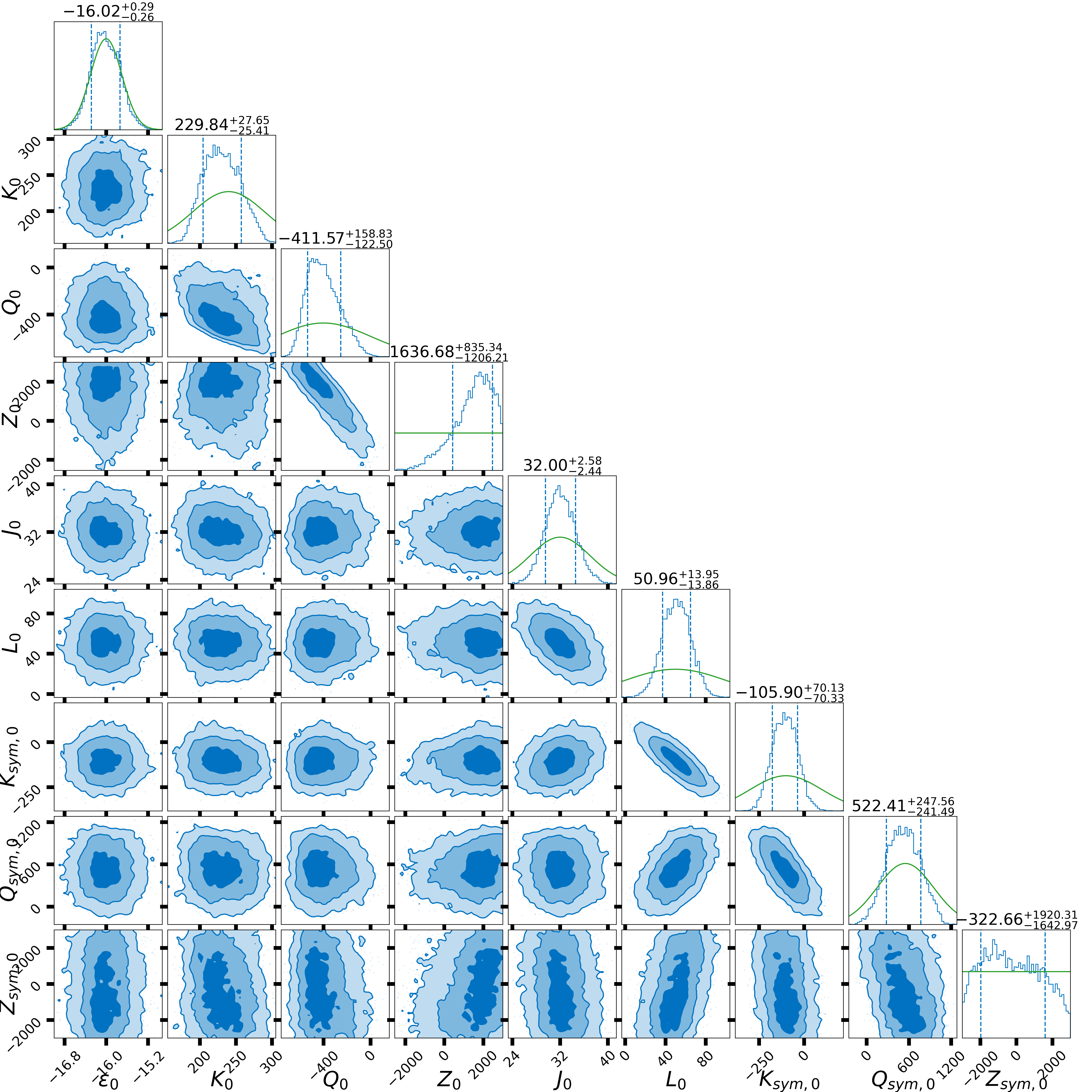}
    \caption{Same as Fig. \ref{fig4}, but, for the model M2.}
    \label{fig5}
\end{figure}
It may be pointed out that there are several additional sources of
uncertainties on the NMIt is customary
to study different correlations in the posterior involving parameters
and observables alike, within a Bayesian analysis. …Ps  which were avoided  
by reconstructing them  separately from the
$\varepsilon(\rho,0)$ and $J(\rho)$ as shown in Figs. \ref{fig1} and \ref{fig2}. 
These sources of uncertainties are (i) inter-correlations
of NMPs corresponding to   $\varepsilon(\rho,0)$ with the ones for   $J(\rho)$, (ii) compensation in the change of 
 $J(\rho)$ with the asymmetry parameter $\delta$ and  $\varepsilon(\rho,0)$ in such a way that the EoS of neutron star matter remains more or less unaltered. We analyze them in detail in the following.

The corner plots for the marginalized PDs for the NMPs in one and
two dimensions for the models M1 and M2 obtained for prior set P2 are
displayed in Figs. \ref{fig4} and \ref{fig5}, respectively. The difference
between the one-dimensional PDs for the NMPs and corresponding prior distributions 
reflect the role of pseudo data in constraining the NMPs. These marginalized posterior distributions of the NMPs are at variance with those obtained
separately from the EoS for the symmetric nuclear matter and the density-dependent symmetry energy as shown in Figs. \ref{fig1} and \ref{fig2}. The
shapes and the orientations of the confidence ellipses suggest that the
correlations among most of the pairs of NMPs have disappeared or weakened. Strong correlations exist only between   $K_0 - Q_0$ , $Q_0 -Z_0$ and
$L_0 - K_{\rm sym,0}$ pairs with correlation coefficient $r \sim
0.8 $  for model M1.  However, in model M2 the $K_0-Q_0$ correlation got disappeared. The inter-correlations of the NMPs corresponding to
$\varepsilon(\rho,0)$ with those for the $J(\rho)$ are almost absent. $Z_0 $ and $Z_{\rm sym,0}$ show almost no correlation with the remaining NMPs.
Overall reduction
in the correlations among the NMPs which are reconstructed from the EoS of the neutron star
matter, but,  increase in their uncertainties at the same time seem to be 
somewhat counter-intuitive. Other sources of uncertainties as mentioned
earlier need to be addressed.

\begin{figure}[]
    \centering
    \includegraphics[width=0.5\textwidth,height=0.3\textheight]{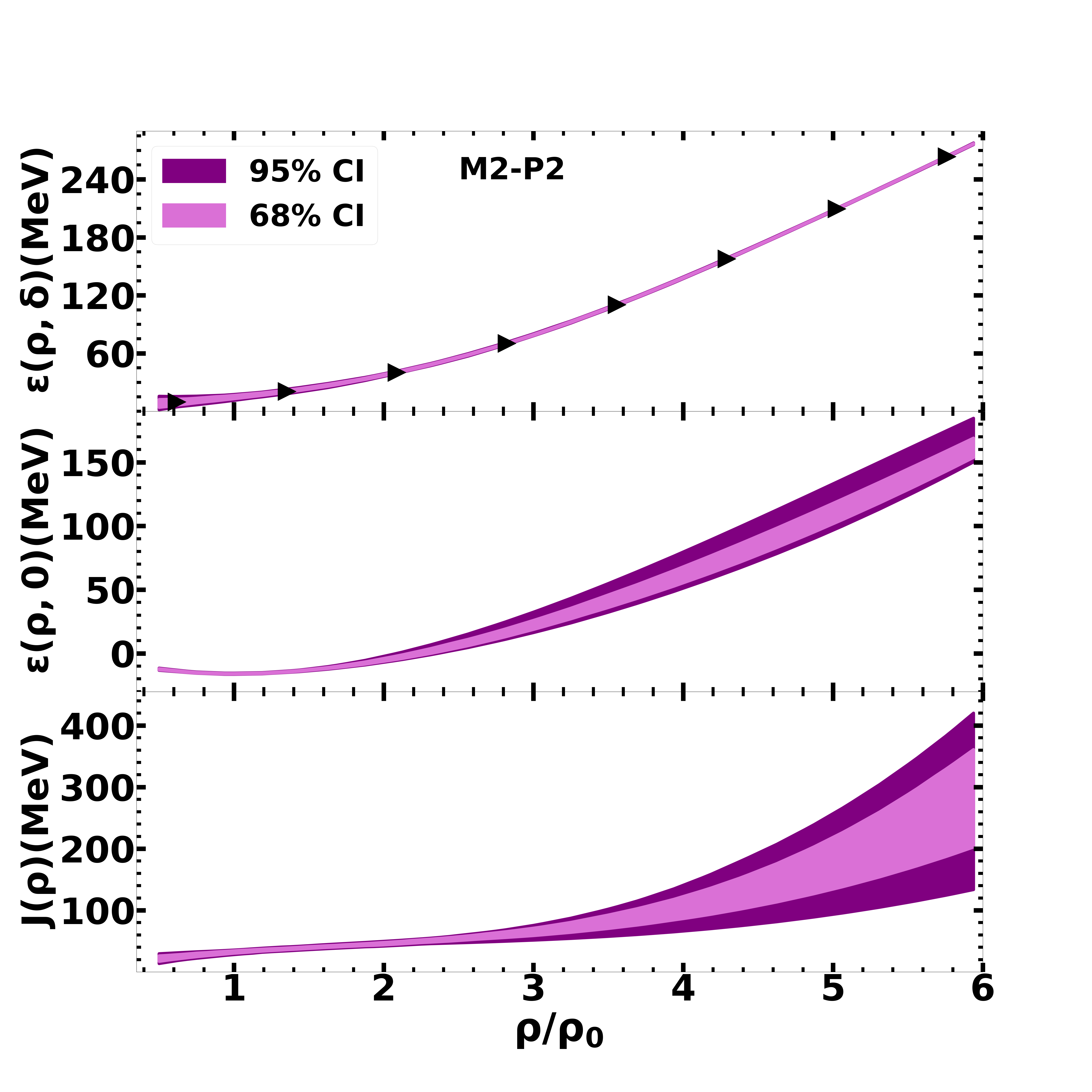}
    \caption{
Plots of 68\% and  95\% confidence intervals for the
EoS for neutron star matter (top),  the symmetric nuclear matter (middle), and the symmetry energy (bottom)  as a function of scaled density for model M2 with prior set P2.
The results are obtained from the posterior distributions of the NMPs
which are reconstructed from the pseudo data for the  EoS of neutron
star matter( triangles).  The spread in  $\varepsilon(\rho,0)$ and  $J(\rho)$  are
consistent with those for  $\varepsilon(\rho,\delta)$.}
\label{fig6} 
\end{figure}

We now examine  the uncertainties in the NMPs which  might arise due
to the allowed variations  in the $\varepsilon(\rho,0)$, $J(\rho)$ and $\delta$ 
for a given $\varepsilon(\rho,\delta)$.
The value of asymmetry parameter $\delta$ is mainly governed by the
symmetry energy at a given density. As the symmetry energy increases,
the $\delta$ decreases.  Thus, the symmetry energy and $\delta$ may
balance each other in such a way that the asymmetric part of the EoS of
neutron star matter remains unaffected.  Moreover, the variations in the
asymmetric part of  $\varepsilon(\rho,\delta)$ may also be compensated
by the symmetric nuclear matter $\varepsilon(\rho,0)$.
In short, for a given $\epsilon(\rho,\delta)$,  the values of $J(\rho)$,
$\varepsilon(\rho,0)$ and $\delta$ may have some leeway.
We use the marginalized PDs for the NMPs to obtain 68\% and
95\% confidence intervals for $\varepsilon(\rho,\delta)$, $\varepsilon(\rho,0)$ and $J(\rho)$ 
. The results are plotted only for the M2-P2
case in Fig. \ref{fig6}. Other cases show similar qualitative trends and are not
shown here.
The  value of $\varepsilon(\rho,\delta)$ (top)  vary in a narrow bound at a given density, but,  $\varepsilon(\rho,0)$ (middle) and $J(\rho)$ (bottom)  have larger uncertainties. The $95\%$ confidence intervals for $\varepsilon(\rho,0)$ and $J(\rho)$
are little asymmetric  with respect to the ones for the $68\%$
 due to the non-Gaussian nature of higher-order NMPs as can be seen
from Table \ref{tab4} and  Fig.  \ref{fig5}. The spread in the values of $J(\rho)$  increases
with density rapidly beyond 2$\rho_0$. For  $68\%$ confidence interval, spread in  
$J(\rho)$ at 4$\rho_0$ is $\sim$ 36 MeV which increases to $\sim$ 160 MeV  at 6$\rho_0$,
whereas, the spread in $\varepsilon(\rho,0)$  remains almost the same ($\sim$ 15 MeV) for the density in the range 4$\rho_0$ to 6$\rho_0$. The larger spread in $J(\rho)$ is balanced by
asymmetry parameter $\delta$ as well as by change in $\varepsilon(\rho,0)$
such that the EoS for neutron star matter remains almost unaffected.
\begin{figure}
    \centering
\includegraphics[width=0.5\textwidth,,height=0.3\textheight]{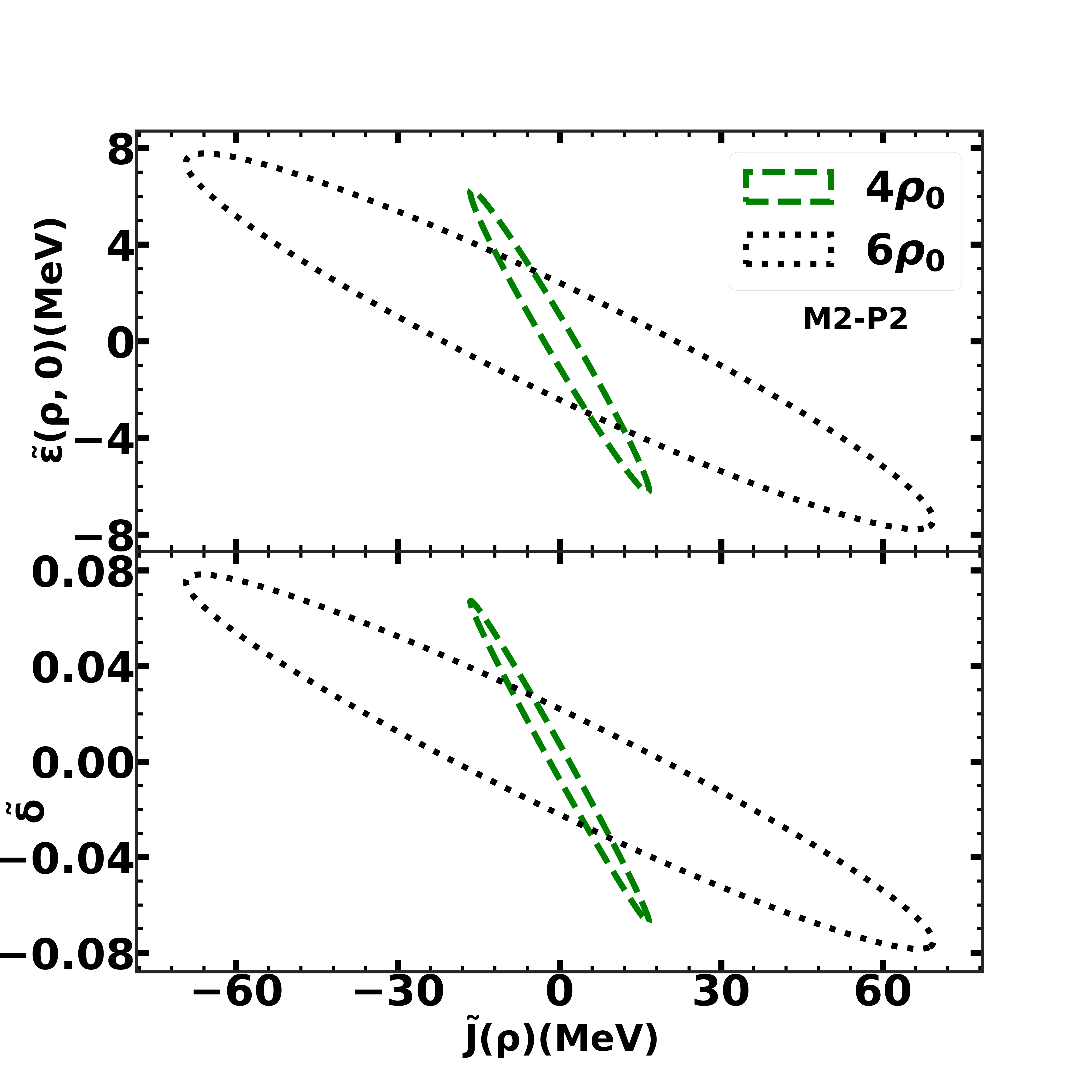}
    \caption{
Plots of  confidence ellipses with $1\sigma$ interval  for the EoS for
symmetric nuclear matter (top) and asymmetry parameter (bottom) as a
function of  symmetry energy at densities $\rho = 4\rho_0$ and $6\rho_0$.
The symbol tilde denotes that the corresponding quantity is obtained
with respect to its median value. }\label{fig7}
\end{figure}
These features are intrinsic in nature which is present in all the
models for neutron star matter (cf. Eq. \ref{eq:eos}).  
 The marginalized PDs
for the NMPs,  plotted in Figs. \ref{fig4} and \ref{fig5}  effectively
correspond to the values of $\varepsilon(\rho,0)$  and $J(\rho)$
displayed in Fig. \ref{fig6}.  To probe further, the confidence ellipses  are plotted in
Fig. \ref{fig7} for the  $68\%$ confidence intervals for $\varepsilon(\rho,0)$,
and $\delta$ as a function of $J(\rho)$ at fixed densities $\rho = $ 4$\rho_0$
and  6$\rho_0$. The $J(\rho)$ is anti-correlated  with $\delta$
and $\varepsilon(\rho,0)$.  The spread in the values of $J(\rho)$
is predominantly  due to its anti-correlation with $\delta$. The uncertainties in $\varepsilon(\rho,0)$ and $J(\rho)$  propagate into the NMPs. That is why, the marginalized posterior  distributions of NMPs   displayed in Figs. \ref{fig4} and \ref{fig5} are significantly  different  in comparison  to those shown in Figs. \ref{fig1} and \ref{fig2}. This also 
explains the reason behind  the larger  uncertainties on the higher-order NMPs, which govern the high-density behavior of $J(\rho)$. As the uncertainties
in $\varepsilon(\rho,0)$ are smaller than those in $J(\rho)$, they get reflected in the  uncertainties of the corresponding NMPs  (see Table \ref{tab4}).  
It seems that the EoS of neutron star matter, usually constrained by using several astrophysical observables alone may not be sufficient to determine the NMPs in narrow bounds. The more reliable determination would also require additional constraints on the EoS of symmetric nuclear matter as well as on the density-dependent symmetry energy. The experimental data on the EoS of symmetric nuclear matter from the heavy-ion collision and the symmetry energy beyond the saturation density from the isobaric analog states may help in constraining the NMPs further  \cite{Danielewicz:2002pu,Tsang:2008fd,Danielewicz:2013upa,Horowitz:2014bja,Roca-Maza:2018ujj}.

\begin{table}[htbp]
	\caption{ Values of correlation coefficients among selected pairs of nuclear matter parameters obtained with the modified prior set P2$^\prime$. The width of Gaussian priors for $Q_0$ and $L_0$  are reduced by factor of four and that for $K_{sym,0}$ by factor of two  for P2$^\prime$ in comparison to the prior set P2 }
	\label{tab5}
	\centering
	\begin{ruledtabular}  
		\begin{tabular}{ccccc}
			Parameters&M1-P2& M1-P2$^\prime$ &M2-P2&M2-P2$^\prime$\\ [1.5ex]
			\hline 
			$K_0-Q_0$ & -0.9 & -0.75 & -0.5  & -0.15   \\[1.5ex]  
		
			$Q_0 -Z_0$ &- 0.97 & -0.87 & -0.88      &  -0.63  \\[1.5ex]  
			{$K_{\rm sym,0}-L_0$} &-0.76 & -0.44 & -0.86   & -0.62 \\[1.5ex]  

		\end{tabular}
	\end{ruledtabular}
\end{table}

\begin{table}[htbp]
\caption{ Same as \ref{tab4}, but, for a modified priors set P2$^\prime$.  \label{tab6}}

 \centering
 \begin{ruledtabular}  
\begin{tabular}{cccccc}
   NMPs & M1-P2$^\prime$ & M2-P2$^\prime$ & NMPs & M1-P2$^\prime$& M2-P2$^\prime$\\ [1.5ex]
\hline 
{$\varepsilon_0$} &  $-16.0_{-0.3}^{+0.3}$  &  $-16.0_{-0.3}^{+0.3}$& {$J_0$}   & $32.0_{-2.0}^{+2.0}$     &  $32.0_{-2.1}^{+2.2} $  \\[1.5ex] 
 & & & {$L_0$} & $51.6_{-7.5}^{+7.4}$  & $50.0_{-6.9}^{+7.2}$     \\[1.5ex]
{$K_0$} & $212_{-21}^{+24}$    & $225_{-21}^{+27}$ &{$K_{\rm sym,0}$}  &  $-87_{-44}^{+45}$ & $-105_{-41}^{+43}$     \\[1.5ex]  

{$Q_0$}  & $-425_{-55}^{+63}$      &  $-408_{-59}^{+62} $ &{$Q_{\rm sym,0}$}  & $489_{-222}^{+219} $   & $544_{-209}^{+213}$    \\[1.5ex]  
    
{$Z_0$}  & $1561_{-80}^{+75}$   & $1650_{-647}^{+573}$ & {$Z_{\rm sym,0}$}  &$ -96_{-604}^{+775}$    & $-332_{-1621}^{+1857}$    \\[1.5ex]

  
\end{tabular}
\end{ruledtabular}
\end{table}
We modify the  prior distributions to simulate the influence of the constraints on the NMPs derived from
the data on the microscopic systems such as heavy-ion collisions and
the bulk properties of finite nuclei. These data are expected to constraint the behavior of symmetric nuclear matter and symmetry energy over a wide range of densities ranging from sub saturation density to supra saturation densities up to 2-3$\rho_0$. The empirical values of pressure of the symmetric nuclear matter at supra saturation densities may constrain the value of $Q_0$. The data on iso-vector giant dipole resonance and neutron-skin thickness in heavy nuclei may constraint the value of $L_0$ \cite{Roca-Maza:2018ujj}. Once the values of $J_0$ and $L_0$ are constrained, $K_{\rm sym,0}$ may also be somewhat constrained \cite{Tews2017,Mondal2017u}. 
We repeat our calculations by reducing the width of Gaussian priors
 for $Q_0$, $L_0$ and $K_{\rm sym,0}$ in the prior set P2. For
the $Q_0$ and $L_0$, the values of width are reduced by a
factor of four, whereas, for the $K_{\rm sym,0}$ by a factor of two. In
Table \ref{tab5}, we present the values of correlation coefficients
among some selected pairs of NMPs obtained with the modified priors with those for the prior set P2. In general, the correlations become weaker with the modified prior. Consequently,
the uncertainties on the NMPs have decreased as can be seen from Table \ref{tab6}.  In particular, the uncertainties on $Z_0$ has now become almost half   for both the models
M1 and M2.  The spread in the values of $\varepsilon(\rho,0)$ and $J(\rho)$ become
smaller by less than $10\%$ for the densities around  $2 \rho_0$. But,
their spreads at higher densities remain practically unaltered. The issue presented in this paper needs further investigation. In the present work, we have used the most commonly employed EoS expanded around the symmetric nuclear matter. Some alternative representation of the EoS of neutron star matter may be employed. One such form is the expansion of the EoS around the neutron matter in powers of the proton fraction \cite{Forbes2019}.



\section{Summary and outlook} A Bayesian approach has been applied to reconstruct
the underlying nuclear matter parameters which describe the EoS of
the neutron star matter. The calculations are performed using the EoS for neutron star matter by expanding it around symmetric nuclear matter within the parabolic expansion as commonly employed. The EoS of symmetric nuclear matter and density-dependent symmetry energy required for such an EoS are expanded using Taylor and ${\frac{n}{3}}$ expansions.  The expansion coefficients for the former are the individual nuclear matter parameters and the linear combinations of them for the ${\frac{n}{3}}$ case.  The pseudo data for the EoS for symmetric nuclear matter, neutron star matter, and density-dependent symmetry energy are constructed using both expansions. This pseudo data enable us to identify the various sources of uncertainties associated with the marginalized posterior distributions of NMPs, since, the true models are known.   The posterior distributions
of the nuclear matter parameters are obtained using two different sets
of priors. One of the prior sets assumes that most of the parameters are
unknown, except for the lowest order ones which  are the binding energy
per nucleon for the symmetric nuclear matter and the symmetry energy
coefficient at the saturation density.

The marginalized posterior distributions for the NMPs
reconstructed separately from the EoS of symmetric nuclear matter and
density-dependent symmetry energy are very much localized around their
true values. But, the posterior distributions for all the NMPs determined simultaneously from the EoS of neutron star matter are at variance.  The median values significantly deviate from their true values and associated uncertainties are also larger, in particular, for second or higher-order NMPs. The main
sources of uncertainties are found to be (i) the correlations among
higher-order parameters describing the EoS of symmetric nuclear matter
and similar correlations in the case of  density-dependent symmetry
energy and (ii) the larger uncertainties in the symmetry energy at a given density due to its anti-correlation with asymmetry parameter and the  EoS of symmetric nuclear matter such that neutron star matter EoS remains mostly unaffected. These are intrinsic in nature for the EoS of neutron star matter obtained by expanding it around the symmetric nuclear matter. 
The EoS of neutron star matter alone may not be sufficient to determine
the higher-order NMPs in narrow bounds. The higher-order  NMPs are correlated to the lower-order ones, thus, the low-density ab-initio predictions for the EoS of symmetric nuclear matter and pure neutron matter from the chiral effective field theory should also be considered for the improved parameterizations.   The experimental data on the EoS of symmetric nuclear matter from the heavy-ion collision and the symmetry energy beyond the saturation density from the isobaric analog states may further help in constraining the nuclear matter parameters. 
 
We have also performed the calculations by imposing stringent constraints on prior distributions for $Q_0$, $L_0$ and $K_{sym,0}$ which led to the reduction of the correlations among the NMPs. Consequently, the uncertainties on some of the NMPs become smaller. The spread in the EoS of symmetric nuclear matter and symmetry energy become
smaller by less than $10\%$ for the densities around  $2 \rho_0$. But,
their spreads at higher densities remain practically unaltered. It remains to be understood whether the sources of various uncertainties identified in the present work are due to the expansion of the EoS around the symmetric nuclear matter.  It may be interesting to perform an investigation using the EoS expanded around the pure neutron matter instead of symmetric nuclear matter\cite{Forbes2019}.

 \section{Acknowledgements} The authors would like to thank Arunava
 Mukherjee, Saha Institute of Nuclear Physics for useful discussion,
 a careful reading of the paper, and important suggestions. N.K.P.
 would like to thank T.K. Jha for constant encouragement and support and gratefully acknowledge the Department of Science and
 Technology, India, for the support of DST/INSPIRE Fellowship/2019/IF190058. C.M. acknowledges partial support from the IN2P3 Master Project “NewMAC”.
\newpage
 \bibliographystyle{apsrev4-1}
  \bibliography{library}

\end{document}